# Multi-Objective Channel Allocation in Cognitive Radio Networks

Ahmad Ghasemi

Worcester Polytechnic Institute

ahmad.ghasemi@mailfence.com

Foad Ghasemi

Regional Water Company of Hormozgan

foadghasemi@mailfence.com

**Abstract** — This paper addresses the problem of channel allocation in Cognitive Radio (CR) networks. CR has been considered as a technology which improves spectrum utilization significantly by carrying out Dynamic Spectrum Management (DSM). One issue of the DSM is the using of frequency channels by secondary/CR users that are under-utilized and/or not used by primary users. CR users eager to use them when they are not used by primary users. The number of these spectrum bands is limited. This scarcity leads to conflict among CR users; As many as bands used by any CR user leads to decaying objective functions of other users. Thus, they have a destructive effect on each other. The paper models this conflict as a new multi-objective problem. Pareto set is attained via a Multi-Objective Optimization (MOO) technique, namely ε - constraint method. Results show the efficiency of the method.

## I. INTRODUCTION

The demand for wireless spectrum use has been growing dramatically. This results in scarcity in the available spectrum bands. A lot of allocated frequency bands are under-utilized temporally and/or geographically [1]. For instance, only 4.8% of the radio spectrum were in use in the United States in 2009 [2]. One approach to utilize these resources, in a more efficient way is dynamic spectrum access (DSA). That also is referred as cognitive radio (CR). CR is capable to sense, learn, and adapt to the outside world. It includes a spectrum sensing, spectrum management, spectrum mobility, and spectrum sharing [3].

There are two types of users in a network; primary and secondary/CR users. Primary users specifically devoted frequency bands. That's not the case for CR users. They sense the spectrum and find available bands to use for communication. Here, the problem is how to allocate the available channels/bands to CR users efficiently. There is a lot of research for CR channel allocation in the literature. Based on cooperative /non-cooperative channel allocation behavior, centralized/decentralized architecture, different methods such as game theory [4], pricing and auction mechanisms [5]-[6], and graph coloring [2,7,8] have been proposed for DSA. Author in [9] model the spectrum sharing among multiple SUs and only one primary user (PU) as an oligopoly market competition. They use a non-cooperative game to execute the channel allocation among SUs. Adaptive allocation channels and transmission power among SUs based on ambient status without disturbing Pus is considered in [10]. Authors in [2] discuss an opportunistic spectrum-sharing. They show that DSM is equivalent to a graph-coloring problem (GCP). Reference [11] considers distributed channel allocation for OFDM based systems for fully connected networks. CR users purchase channels from a PU through an auction process in [5]. The proposed payment metric is based on receiving signal-to-noise ratio. Authors in [12] propose a dynamic pricing approach to optimize overall spectrum efficiency while keeping the participating incentives of the users based on double auction rules. A survey on Centralized DSA is presented in [13]. In most of the previous works, spectrum allocation chiefly was modeled as a single-objective problem [14]-[16].

In CR networks, each CR user wants to enhance its reward/utility. This leads to decay the other user's rewards/utilities. It means that utilities have destructive effects on each other. Thus, this paper model the spectrum allocation as a new multi-objective problem in which CR utilities are the objectives. This channel allocation is such that the created interference to the PUs by CR users, and the produced interference among SUs is minimized while our goal is to fairly distribute the channels among CR users. The paper solves the problem using a MOO technique, namely $\varepsilon$ - *constraint* method. Corresponding challenges from technical points of view are investigated.

In this paper, capital and small boldface letters show matrices and vectors, respectively. The rest of the paper is organized as follows. Access to the spectrum of CR based on open spectrum systems is described in Section II. In Section III, a mathematical model of open spectrum access is provided, and MOO problem is defined. In Section IV, a technique for solving the defined problem, i.e. $\varepsilon$ - constraint method is presented in detail. Simulation results are provided in Section V, and finally, Section VI concludes the paper.

## II. NETWORK MODEL

We use the network model in [8] that is in the context of open spectrum systems such as CR networks. CR users use unused/underutilized licensed spectrum bands. The goal is to maximize utilization, but they must act in a non-interfering manner based on imposing constraints by PUs. Note that CR users automatically detect footprints of PUs via accessing a central database. PUs can be different access points in a mesh network or base stations serving for different wireless network operators.

Here, the model explores the network and provides available channels and interfering SUs for all active SUs in the network as explained in the following example. There is one PU **A** in channel $m$. Its *protection area* is a circle with the radius $d_p(A, m)$ (see Fig. 1). In addition, four SUs exist in the network that can communicate data on the channel. Each SU should regulate its circular *interference range* to prevent any interference effect on PU. $d_s(i, m)$ denotes the radius of the interference range of the $i^{th}$ SU. An upper bound $d_{\max}$ of SUs' interference range is limited to the border of PU's protection area to avoid interfering with PU. $d_{\max}$ corresponds to the maximum transmit power, i.e., $P_{t_{\max}}$. Moreover, $d_{\min}$ of an SU interference range corresponds to the minimum transmit power, i.e., $P_{t_{\min}}$. $d_{\min}$ for a given channel is determined by the minimum of two values: (1) the minimum distance to all the PUs present on that channel, and (2) the distance corresponding to its maximum transmitting power. Thus, SUs can't exploit the channel if they are located inside of the protection area such as the $4^{th}$ SU in Fig. 1.

Interference among SUs must be determined after determining their interference ranges for all channels. The relative position of SUs is interpreted by $dist(i,j,k)$ which is the distance between the interference range centers of the $i^{th}$ and $j^{th}$ SUs in channel $k$. Two users would interfere with each other if their corresponding relative position be less than or equal to the sum of the radiuses of their interference ranges, i.e., $dist(i,j,k) \leq d_s(i,k) + d_s(j,k)$ such as SUs 2 and 3 in Fig. 1.

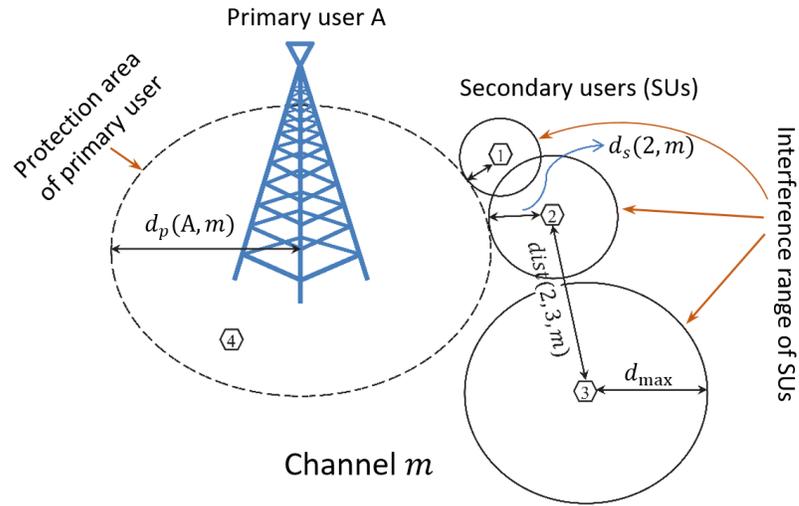

Fig. 1. Availability of a given channel for the primary and secondary users.

### III. ALLOCATION MODEL AND UTILITY FUNCTIONS

In this section, we explain a theoretical model to indicate the general allocation problem; Next, a new multi-objective problem is formulated. Here, we assume a slow varying spectrum environment where users quickly conform to environmental changes and their location don't change during the allocation period.

There are $N$ SUs competing for $M$ available spectrum channels. The channel availability for each SU are determined by the location of nearby PUs and their channel usage. The key components of the model are *channel availability*, *channel reward*, *Interference constraint*, *conflict-free channel allocation*, *radio interface limit*, and *user reward*. Appendix I presents a pseudo-code for generating the network model and its components.

*Channel availability* **L** is a $N \times M$ binary matrix where its $(n,m)^{th}$ element is one, i.e., $[\mathbf{L}]_{n,m} = 1$, only if channel $m$ is available for the $n^{th}$ user. In addition, if $d_s(n,m) < d_{\min}$ then $[\mathbf{L}]_{n,m} = 0$, otherwise $[\mathbf{L}]_{n,m} = 1$.

*Channel reward* **B** is a $N \times M$ matrix where its $(n,m)^{th}$ element, i.e., $[\mathbf{B}]_{n,m}$, is defined in (1), and denotes the maximum throughput/coverage that can be exploited by the $n^{th}$ user at channel $m$.

$$[\mathbf{B}]_{n,m} = \begin{cases} d_s(n,m)^2 & \text{iff channel } m \text{ is avaiable for the } n\text{th user} \\ 0 & \text{if } [\mathbf{L}]_{n,m} = 0 \end{cases}, \text{ where } d_{\min} \leq d_s(n,m) \leq d_{\max} \quad (1)$$

*Interference constraint* **C** is a $N \times N \times M$ binary matrix. The $(n, k, m)^{th}$ element of this 3D matrix, i.e., $[\mathbf{C}]_{n,k,m}$, presents the interference between $n^{th}$ and $k^{th}$ users at channel $m$. Its value is one only if these users interfere at that channel and they are located within a certain distance, i.e. $dist(n, k, m) \leq d_s(n, m) + d_s(k, m)$.

Note that the interfering is a channel specific feature; two users may interfere on one channel but not another.

*Conflict-free channel allocation* **A** is a $N \times M$ binary matrix where its $(n,m)^{th}$ element, i.e., $[\mathbf{A}]_{n,m}$, is one if channel $m$ is assigned to the $n^{th}$ user. This matrix should satisfy the interference constraints defined by **C** that is represented in (2).

$$[\mathbf{A}]_{n,m} + [\mathbf{A}]_{k,m} \leq 1, \quad \text{if } [\mathbf{C}]_{n,k,m} = 1, \ \forall\, n, k < N, \ m < M, \quad (2)$$

*Radio interface limit* $C_{\max}$ denotes the maximum number of channels that is assigned to an SU. The channel allocation for each user $n$ satisfies $\sum_{m=1}^{M}[\mathbf{A}]_{n,m} \leq C_{\max}$.

Based on the provided definitions, the reward for the $n^{th}$ user, i.e., $[\mathbf{r}]_n$, is a vector with length $N$ that is defined as $[\mathbf{r}]_n = \sum_{m=1}^{M}[\mathbf{A}]_{n,m}[\mathbf{B}]_{n,m}$.

Here, we define an optimization function in (3) in which $U(\mathbf{r})$ is a utility function based on $\mathbf{r}$.

$$\mathbf{A}^* = \arg\max_{\mathbf{A}} \ U(\mathbf{r}) \quad (3)$$

In the sequel, we formulate (3) as a *multi-objective* problem and its optimal solution is denoted by a binary matrix $\mathbf{A}^*$. Some definition should be explained before the problem formulation.

A multi-objective mathematical programming (MMP) includes at least two objective functions (OF) in which there is no single optimal solution that simultaneously optimizes all OFs. Here, the concept of optimality is replaced with that of *Pareto optimality*. A Pareto optimal (or non-inferior/non-dominated) is a solution that cannot be improved for one OF without declining the performance of at least one of other OFs. Thus, the solution is to find a set of Pareto optimal, i.e., *Pareto set*.

We provide the multi-objective optimization problem in (4). Here, each SU has own OF. This problem is a mixed integer programming because the linear constraints, integer elements in of matrices **A** and **C**, and the non-negative elements of matrix **B**.

$$\begin{aligned} \arg\max_{\mathbf{A}} \quad & [\boldsymbol{r}]_1, \ldots, [\boldsymbol{r}]_N \\ \text{s.t.} \quad & [\mathbf{A}]_{n,m} + [\mathbf{A}]_{k,m} \leq 1 \quad \text{if } [\mathbf{C}]_{n,k,m} = 1 \quad \forall 1 \leq n, k \leq N, \ 1 \leq m \leq M \\ & \sum_{m=1}^{M} [\mathbf{A}]_{n,m} \leq C_{max} \quad , [\mathbf{A}]_{n,m} \in \{0,1\} \end{aligned} \quad (4)$$

## IV. PROBLEM SOLVING

There are different approaches in the literature to solve a multi-objective problem. One approach is *weighted sum* method [17] in which the set of objectives are scalarized into a single objective by multiplying each objective with a user-supplied weight. This simple method suffers from some disadvantages such as a uniformly distributed set of weights does not guarantee a uniformly distributed set of Pareto optimal, and two different set of weight vectors doesn't necessarily lead to two different Pareto sets. Thus, the paper uses *ε-constraint* method [18] to reach the Pareto set of the optimization problem (4). ε-constraint method has several advantages over the weighted sum method [18]: I) The scaling of the OFs is not necessary for this method; and II) The number of the generated efficient solutions can be controlled. We explain this method in the following sub-sections.

### A. *Solution methodology:*

The ε-constraint method considers one of OFs as the main objective to optimize and the others as constraints. If the first objective function is considered as the main OF, the formulation of the multi-objective optimization problem is as follows [18]:

$$\begin{aligned} \min/\max_{\mathbf{X}} \quad & \left[ f_1(\mathbf{x}) + d_1 \varepsilon \sum_{i=2}^{p} s_i / r_i \right] \\ \text{s.t.} \quad & f_i(\mathbf{x}) - d_i s_i = e_{i,n_i}, \ s_i \in R^+, \ i = 2, \ldots, p, \ ni = 0, 1, \ldots, q_i \end{aligned} \quad (5)$$

Here, $\mathbf{x}$ and $p$ refer to the vector of decision variables and the number of OFs, respectively. $d_i$ denotes the $i^{th}$ OF direction that $d_i = -1$ and $+1$ are for minimizing and maximizing the OF. In this paper all OFs should be maximized. Solutions are obtained via *parametrical iterative variations* in $e_{i,ni}$. $s_i$ and $r_i$ are a slack/surplus variable for the constraints, and the range of $i^{th}$ OF, respectively. Here, $s_i/r_i$ is used to avoid any scaling problem. Finally, $\varepsilon$ is a small constant number such as $10^{-6}$.

In order to employ the ε-constraint method, the range of each OF, i.e., $r_i$ should be determined. The common approach is calculating these ranges from the payoff table. The author in [18] proposes a *lexicographic optimization* in order to construct the payoff table including only Pareto optimal. The lexicographic optimization optimizes the first OF and then among the possible alternative optima, optimize the second OF and so on. The details of the computing payoff table are presented in the sub-section B. The payoff table has $p$ rows and columns. The $i^{th}$ column includes the obtained values for the $i^{th}$ OF. The relative difference between the minimum and maximum values of this column indicate the range of the OF, i.e., $r_i = f_i^{max} - f_i^{min}$.

### B. determining payoff table:

Calculating the individual optima of OFs is the first step to determine the payoff table. If $x_i^*$ represents the vector of decision variables that optimizes the $i^{th}$ OF, then the optimum value of this OF is indicated by $f_i^*(x_i^*)$. Next, the value of the other OFs $f_1, f_2, \ldots, f_{i-1}, f_{i+1}, \ldots, f_p$ are determined based on $f_i^*(x_i^*)$, which are represented by $f_1(x_i^*), f_2(x_i^*), \ldots, f_{i-1}(x_i^*), f_{i+1}(x_i^*), \ldots, f_p(x_i^*)$. The $i^{th}$ row of the payoff table includes $f_1(x_i^*), f_2(x_i^*), \ldots, f_i^*(x_i^*), \ldots, f_p(x_i^*)$. All rows are calculated using the same approach. The payoff table is shown in (6).

$$\phi = \begin{bmatrix} f_1^*(x_1^*) & \cdots & f_i^*(x_1^*) & \cdots & f_p^*(x_1^*) \\ \vdots & \ddots & & & \vdots \\ f_1^*(x_i^*) & & f_i^*(x_i^*) & & f_p^*(x_i^*) \\ \vdots & & & \ddots & \vdots \\ f_1^*(x_p^*) & \cdots & f_i^*(x_p^*) & \cdots & f_p^*(x_p^*) \end{bmatrix} \quad (6)$$

Here, we provide some definitions.

*Utopia point* $f^{max}$ is a specific point outside the feasible region, where all objectives are at their best possible values at the same time (see Fig. 2). It is presented as:

$$f^{max} = [f_1^{max}, \ldots, f_i^{max}, \ldots, f_p^{max}] = [f_1^*(x_1^*), \ldots, f_i^*(x_i^*), \ldots, f_p^*(x_p^*)] \quad (7)$$

*Nadir point* $f^N$ is a point in the objective space where all OFs are at their worst values at the same time and is written as $f^N = [f_1^N, \ldots, f_i^N, \ldots, f_p^N]$ in which $f_i^N$ for the feasible region $\Omega$ is defined as:

$$f_i^N = \min_{\mathbf{x}} f_i(\mathbf{x}), \quad \text{s.t.} \ \mathbf{x} \in \Omega \quad (8)$$

*Pseudo nadir* (SN) $f^{min}/f^{SN}$ point is defined as $f^{min} = [f_1^{min}, \ldots, f_i^{min}, \ldots, f_p^{min}]$ in which $f_i^{min} = \min \{f_i^*(x_1^*), \ldots, f_i^*(x_i^*), \ldots, f_i^*(x_p^*)\}$.

Note that utopia, nadir and pseudo nadir points are defined in the objective space that OFs are its dimensions. These points for two objectives are shown in Fig. 2. The range of each objective function in the payoff table is based on utopia and pseudo nadir points as:

$$f_i^{\min} \leq f_i(\mathbf{x}) \leq f_i^{\max} \tag{9}$$

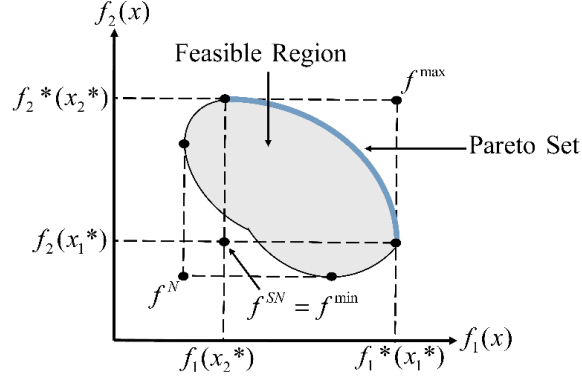

**Fig. 2.** Graphical view of the reference points

Next, ε-constraint technique divides the range of $p - 1$ objective functions $f_2, \ldots, f_p$ obtained via (9) into $n_2, \ldots, n_p$ equal intervals using $(n_2 - 1), \ldots, (n_p - 1)$ intermediate equidistant grid points (GPs), respectively. Considering the minimum and maximum values of the range, there are $(n_2 + 1), \ldots, (n_p + 1)$ GPs for $f_2, \ldots, f_p$, respectively. Thus, we should solve $\prod_{i=2}^{p}(n_i + 1)$ subproblems. The subproblem $(n_2, \ldots, n_p)$ has the following form:

$$\min_{\mathbf{X}} / \max \quad f_1(\mathbf{x}) \tag{10}$$
$$\text{s.t.} \quad f_i(\mathbf{x}) \leq e_{i,n_i}, \quad i = 2, \ldots, p,$$

$$\min \begin{cases} e_{2,n_2} = f_2^{\max} + n_2 \left( \dfrac{f_2^{\min} - f_2^{\max}}{q_2} \right) \\ \vdots \\ e_{p,n_p} = f_p^{\max} + n_p \left( \dfrac{f_p^{\min} - f_p^{\max}}{q_p} \right) \end{cases} \quad \max \begin{cases} e_{2,n_2} = f_2^{\min} + n_2 \left( \dfrac{f_2^{\min} - f_2^{\max}}{q_2} \right) \\ \vdots \\ e_{p,n_p} = f_p^{\min} + n_p \left( \dfrac{f_p^{\min} - f_p^{\max}}{q_p} \right) \end{cases} \tag{11}$$

$$n_2 = 0, 1, \ldots, q_2, \ldots, n_p = 0, 1, \ldots, q_p$$

The outcome of each subproblem is one Pareto optimal. The most preferred Pareto optimal will be selected via a decision maker among the produced Pareto set elements. Note that subproblems with an infeasible solution space will be discarded.

Here, we use a lexicographic optimization to construct the payoff table, including only efficient solutions [18] (see Fig. 3). The lexicographic optimization optimizes the first OF and then among the possible alternative optima optimizes the second OF and so on.

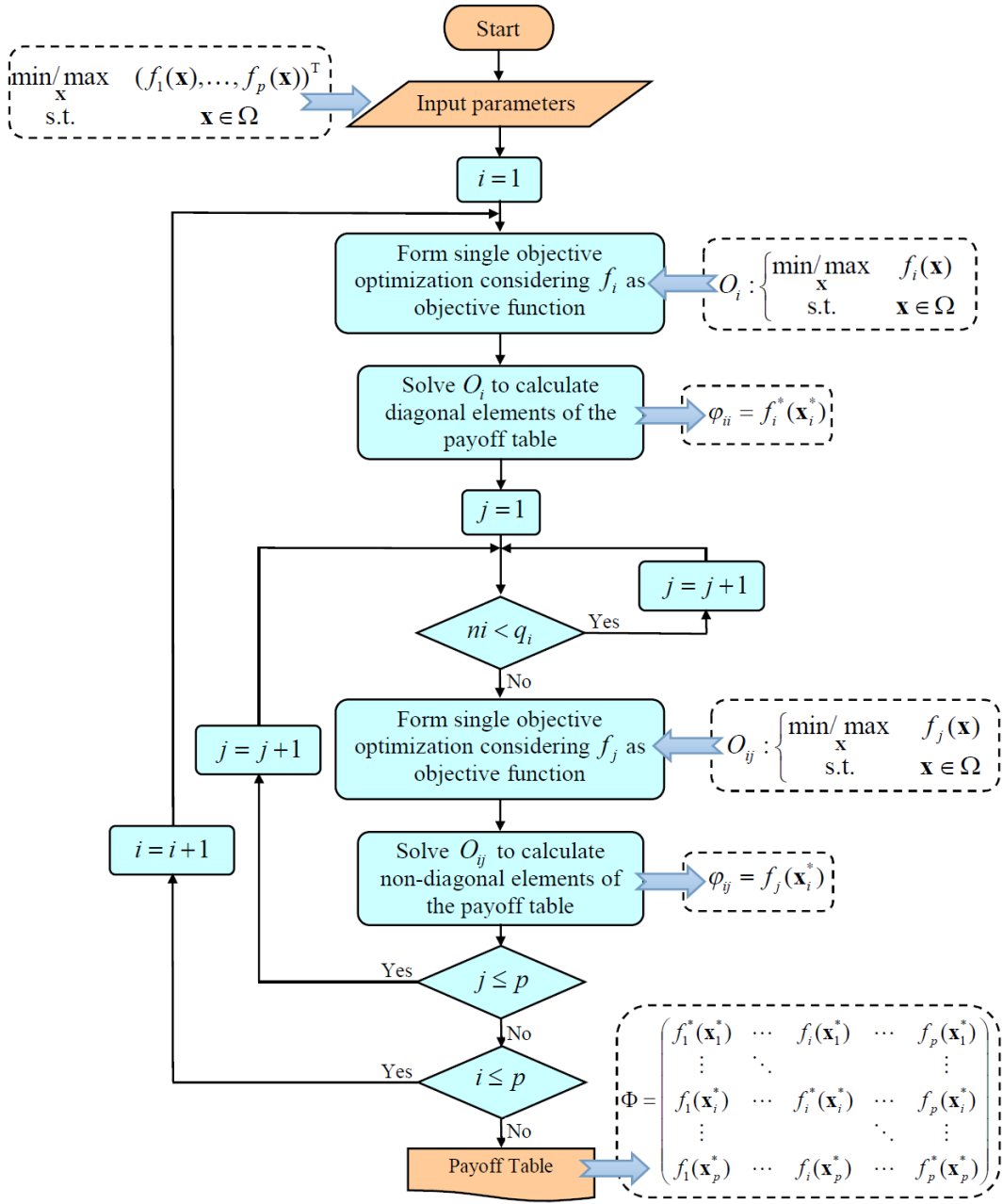

**Fig. 3.** Flowchart of the lexicographic optimization to calculate payoff table

As an example, consider the following optimization problem. The solution space includes the direction of OFs is shown in Fig. 4. The Pareto set is BC ∪ EF.

$$\max \begin{cases} f_1 = 2y \\ f_2 = 4x + 5y \end{cases}$$

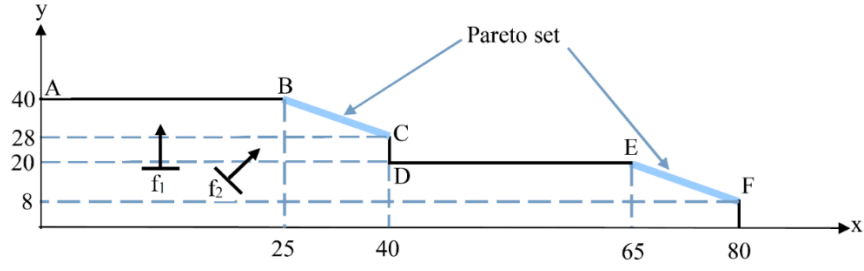

**Fig. 4.** Solution space.

The corresponding payoff table, i.e., $\Phi_1$, is obtained via the lexicographic optimization as:

$$\Phi_1 = \begin{bmatrix} \max f_1 \\ \max f_2 \end{bmatrix} = \begin{bmatrix} 80 & 300 \\ 16 & 360 \end{bmatrix}$$

Here, the obtained solution for max $f_1$ and $f_1$ are correspond to point B and F in Fig. 4, respectively. After calculating the payoff table, the range of remaining OFs should be divided into equal intervals. In this example the remained OF is only $f_2$, i.e., $p - 1 = 1$, and its range is [300, 360]. Assume $q_2 = 8$ equal intervals are obtained. Thus, there are $q_2 + 1 = 9$ GPs. These points are considered as the values for $e_{2,n2}$ ($n2 = 0, 1, \ldots, 8$) in (20). The GPs are $e_{2,0} = 360, \ldots, e_{2,8} = 300$ (see Fig. 5). Fig. 5 shows that the lexicographic optimization determines B, P, Q, R, S, E, T, U, and F as the Pareto set, in which the points B, P, E, U, and F are non-dominated or efficient solutions.

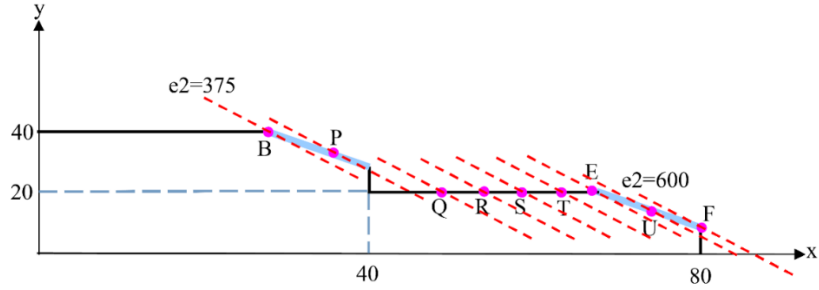

**Fig. 5.** Results of the lexicographic optimization

The proposed procedure of obtaining a Pareto optimal set is depicted in Fig. 6. The proposed method is as follows:

**Step 1:** Computing the payoff table via the lexicographic optimization approach.

**Step 2:** Determining the range of remaining objective functions based on the table, i.e.,

$r_i = f_i^{\max} - f_i^{\min}, i = 2, 3, \ldots, p.$

**Step 3:** Dividing the ranges into $q_i (i = 2, 3, \ldots, p)$ equal intervals via (11).

**Step 4:** Solving the feasible sub-problems in (20) to produce Pareto efficient solutions, whereas the infeasible ones are discarded.

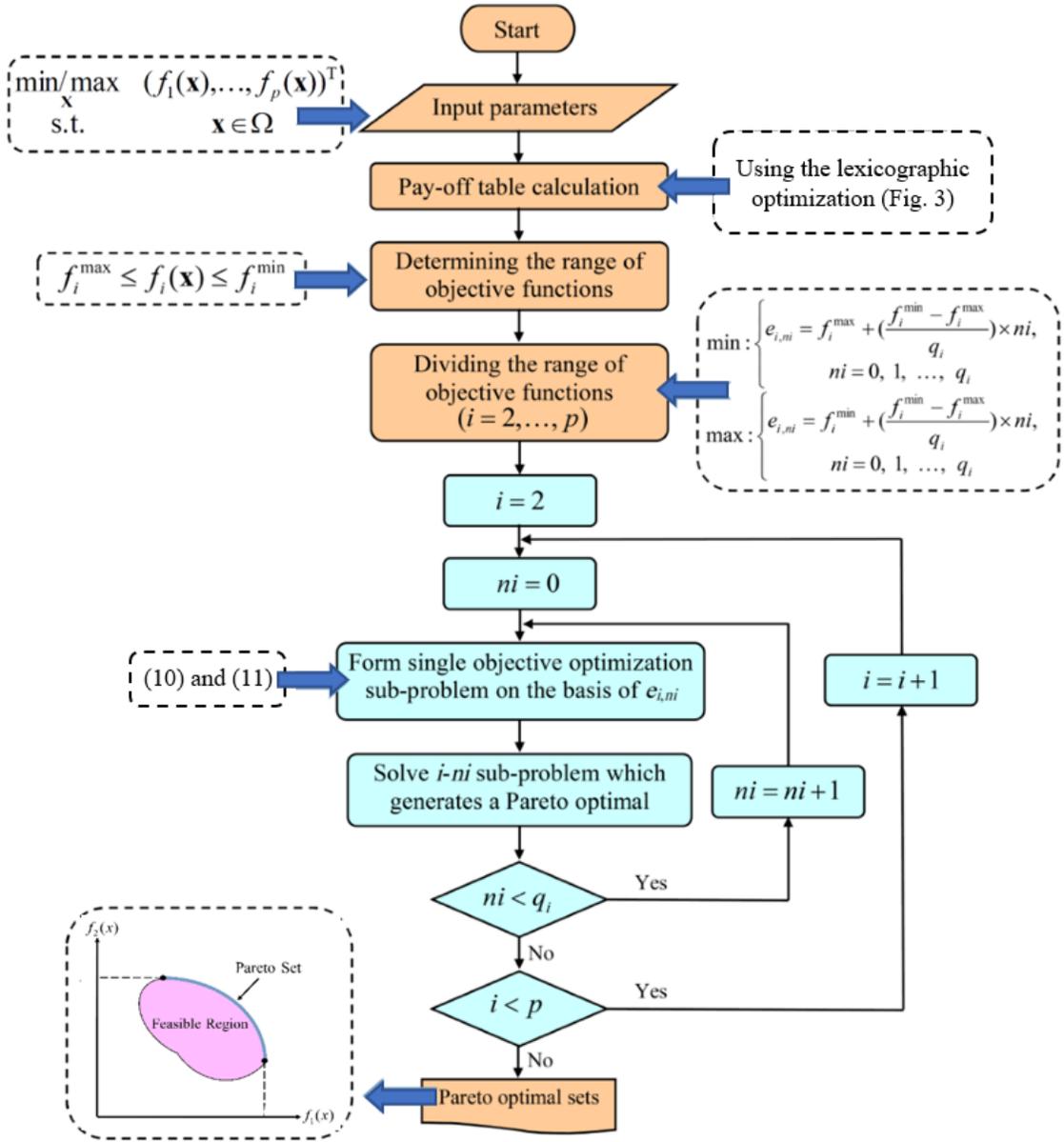

**Fig. 6.** Flowchart of the proposed MMP solution method including augmented ε-constraint with lexicographic optimization.

## V. SIMULATION RESULT

In this section, we evaluate the performance of ε-constraint method to obtain the Pareto set of (4).

### A. Performance Analysis:

Here, a wireless network covering $20 \times 20$ km² square area is considered. We randomly place $K$ PUs and $N$ mobile devices as SUs in the area. In addition, it is assumed that the displacement of the nodes is slower than the convergence time of the proposed algorithm.

We use ε-constraint method coded in the General Algebraic Modeling System (GAMS) [19], using DICOPT solver on a personal computer Intel (R) Core (TM) 2 Due, 2.53 GHz with 4 GB RAM. Here, none of the SUs is superior to the other SUs, thus each OF can be considered as the main OF. We consider the first SUs OF, i.e., $[r]_1$, as the main one. In addition, 20 GPs ($q_i = 20, \ i = 2, 3, \ldots, p$) for $[r]_i, \ i = 2, 3, \ldots, p$ (the other OFs) used for obtaining the Pareto set. In total, to obtain the Pareto set including all feasible solutions, the problem should be solved $q_2 + 1 = 21$ times if $p = 2$, and $(q_2 + 1)(q_3 + 1) = 21 \times 21 = 441$ times if $p = 3$, and so on [18]. Note that it is possible to choose a different number of GPs for each OF.

First, we define two typical multi-objective problems in Table I. The results are shown in Table II and Fig. 7.

**Table I:** The matrices and their optimization problem, a) two SUs, and b) three SUs.

$$[A] = \begin{bmatrix} a_{1,1} & a_{1,2} \\ a_{2,1} & a_{2,2} \end{bmatrix}, \ [B] = \begin{bmatrix} 16 & 16 \\ 16 & 16 \end{bmatrix}, \ [C]_{:,:,1} = \begin{bmatrix} c_{1,1,1} & c_{1,2,1} \\ c_{2,1,1} & c_{2,2,1} \end{bmatrix} = \begin{bmatrix} 0 & 1 \\ 1 & 0 \end{bmatrix}, \ [C]_{:,:,2} = \begin{bmatrix} c_{1,1,2} & c_{1,2,2} \\ c_{2,1,2} & c_{2,2,2} \end{bmatrix} = \begin{bmatrix} 0 & 1 \\ 1 & 0 \end{bmatrix}$$

$$\max_{\mathbf{A}} \begin{cases} [r]_1 = \sum_{m=1}^{2} [A]_{1,m} \cdot [B]_{1,m} = 16a_{1,1} + 16a_{1,2} \\ [r]_2 = \sum_{m=1}^{2} [A]_{2,m} \cdot [B]_{2,m} = 16a_{2,1} + 16a_{2,2} \end{cases}$$

s.t. $[A]_{n,m} + [A]_{k,m} \leq 1$ if $[C]_{n,k,m} = 1 \quad \forall 1 \leq n, k \leq 2, \ 1 \leq m \leq 2$

$$\Rightarrow a_{1,1} + a_{2,1} \leq 1 \text{ and } a_{1,2} + a_{2,2} \leq 1$$

(a)

$$[A] = \begin{bmatrix} a_{1,1} & a_{1,2} & a_{1,3} \\ a_{2,1} & a_{2,2} & a_{2,3} \\ a_{3,1} & a_{3,2} & a_{3,3} \end{bmatrix}, \ [B] = \begin{bmatrix} 16 & 2.0982 & 16 \\ 16 & 16 & 16 \\ 16 & 0 & 16 \end{bmatrix}, \ [C]_{:,:,1} = \begin{bmatrix} c_{1,1,1} & c_{1,2,1} & c_{1,3,1} \\ c_{2,1,1} & c_{2,2,1} & c_{2,3,1} \\ c_{3,1,1} & c_{3,2,1} & c_{3,3,1} \end{bmatrix} = \begin{bmatrix} 0 & 1 & 1 \\ 1 & 0 & 0 \\ 1 & 0 & 0 \end{bmatrix},$$

$$[C]_{:,:,2} = \begin{bmatrix} c_{1,1,2} & c_{1,2,2} & c_{1,3,2} \\ c_{2,1,2} & c_{2,2,2} & c_{2,3,2} \\ c_{3,1,2} & c_{3,2,2} & c_{3,3,2} \end{bmatrix} = \begin{bmatrix} 0 & 0 & 0 \\ 0 & 0 & 0 \\ 0 & 0 & 0 \end{bmatrix}, \ [C]_{:,:,3} = \begin{bmatrix} c_{1,1,3} & c_{1,2,3} & c_{1,3,3} \\ c_{2,1,3} & c_{2,2,3} & c_{2,3,3} \\ c_{3,1,3} & c_{3,2,3} & c_{3,3,3} \end{bmatrix} = \begin{bmatrix} 0 & 1 & 1 \\ 1 & 0 & 0 \\ 1 & 0 & 0 \end{bmatrix}$$

$$\max_{\mathbf{A}} \begin{cases} [r]_1 = \sum_{m=1}^{3} [A]_{1,m} \cdot [B]_{1,m} = 16a_{1,1} + 2.0982 a_{1,2} + 16a_{1,3} \\ [r]_2 = \sum_{m=1}^{3} [A]_{2,m} \cdot [B]_{2,m} = 16a_{2,1} + 16a_{2,2} + 16a_{2,3} \\ [r]_3 = \sum_{m=1}^{3} [A]_{3,m} \cdot [B]_{3,m} = 16a_{3,1} + 16a_{3,3} \end{cases}$$

s.t. $[A]_{n,m} + [A]_{k,m} \leq 1$ if $[C]_{n,k,m} = 1 \quad \forall 1 \leq n, k, m \leq 3$

$$\Rightarrow \begin{cases} a_{1,1} + a_{2,1} \leq 1 \\ a_{1,3} + a_{2,3} \leq 1 \\ a_{1,1} + a_{3,1} \leq 1 \\ a_{1,3} + a_{3,3} \leq 1 \end{cases}$$

(b)

**Table II:** The solutions of Table I

$$\boldsymbol{f}^T = [f_1 \ f_2]^T = \begin{bmatrix} 0 & 16 & 32 \\ 32 & 16 & 0 \end{bmatrix} \quad \boldsymbol{f}^T = [f_1 \ f_2 \ f_3]^T = \begin{bmatrix} 34 & 18.1 & 2.1 \\ 16 & 32 & 48 \\ 0 & 16 & 32 \end{bmatrix}$$

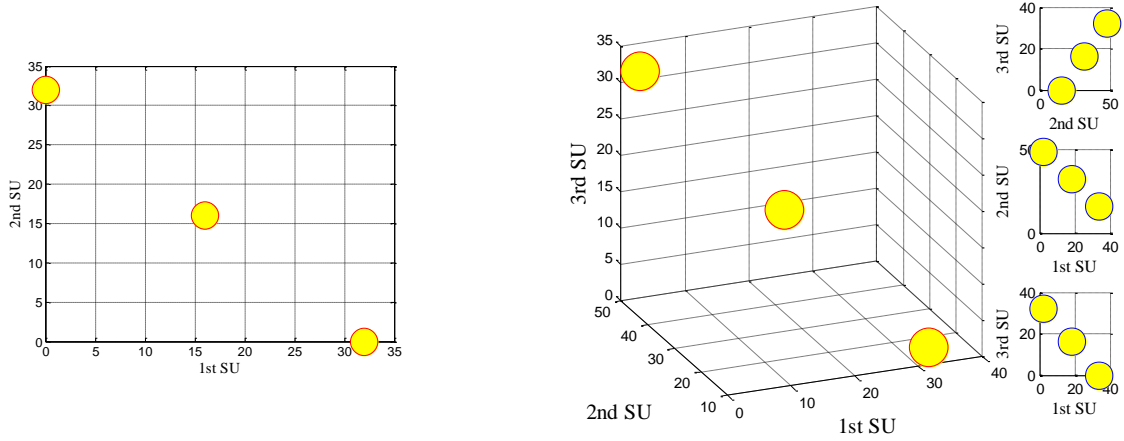

**Fig. 7.** The solution of the problems defined in Table I. Left: two objective functions, Right: three objective functions

Next, we evaluate the algorithm for different topologies with different number of channels $M$, SUs $N$, and PUs $K$. Here, the number of GPs is 10. In the following table, Computation Time (CT) (second), corresponding variance, and standard deviation are shown that they are the averages of 300 different topologies. It can be seen from the table, even though increasing $K$ and $M$ not have significant impacts on the performance of the algorithm but increasing $N$ (i.e. the number of objective functions/SUs) strongly affects the performance. It leads to the CT be very high, i.e., more than 30 hours (h).

**Table III:** The averages computation time, variance, and standard deviation for different topologies.

| $M$ | $N$ | $K$ | Computation Time (s) (CT) | Variance of CT | Standard Deviation of CT | Number of Pareto Optimal Solutions (#POS) | Variance of #POS | Standard Deviation of #POS |
|---|---|---|---|---|---|---|---|---|
| 5 | 5 | 5 | 12.8 | 83.5 | 9.1 | 30 | 1073.1 | 32.8 |
|  |  | 10 | 10.7 | 71.8 | 8.5 | 30 | 798.3 | 28.3 |
|  |  | 15 | 8.2 | 57.2 | 7.6 | 19 | 408.4 | 20.2 |
|  |  | 20 | **5.5** | **6.9** | **2.6** | **15** | **101.1** | **10.1** |
|  | 10 | 5 | 9052.3 | 3.3e+10 | 5787.6 | 9971 | 3.2e+7 | 5721.6 |
|  | 15 |  | > 30 h | -- | -- | -- | -- | -- |
|  | 20 |  | > 30 h | -- | -- | -- | -- | -- |
| 10 | 5 | 5 | 38.9 | 1895.5 | 43.5 | 93 | 15537.9 | 124.7 |
| 15 | 5 |  | 60.6 | 3576.5 | 59.8 | 120 | 11369.3 | 106.6 |
| 20 |  |  | 35.1 | 738.0 | 27.2 | 89 | 7657.5 | 87.5 |

To reach a better performance based on the CT, we can use a lower number of GPs, i.e. less than ten points, but as it was said in [18] accuracy of the algorithm will decline. We present the result for different number of GPs in Table VI. The table shows that the computation time greatly decreases as the number of GPs declines. Thus, there is a balance between accuracy and time based on the number of GPs and determining the optimal number of GPs is critical to reaching the optimal performance for different topologies.

Table VI: The average computation time for two topologies with different GPs.

| M | N | K | GPs | Computation Time (s:second, h:hour) |
|---|---|---|---|---|
| 5 | 15 | 5 | 10 | > 30 h |
|   |    |   | 8  | > 17:31' h |
|   |    |   | 6  | > 7:32' h |
|   |    |   | 4  | 1576.0 s |
|   |    |   | 2  | **24.4** s |
| 5 | 20 | 5 | 10 | > 30 h |
|   |    |   | 8  | > 19' h |
|   |    |   | 6  | > 8 h |
|   |    |   | 4  | > 2:29' h |
|   |    |   | 2  | **336.4** s |

## B. Why ε-constraint?

There are many evolutionary algorithms such as NSGAII [20], MOPSO [21] to solve multi-objective problem and attain a Pareto set. There is no strong guarantee for their convergence. In addition, there are some methods to decrease the number of objective functions. This leads to decreasing the computational complexity of solving procedure. For instance, the authors in [22] propose a method to reduce the number of objective functions via a new mixed-integer linear programming. However, the provided Pareto set via these approaches is not accurate and has some errors. Thus, this paper uses the ε-constraint method to attain a Pareto set.

## VI. CONCLUSION AND FUTURE WORK

The paper formulates a new model as a multi-objective problem for optimizing utilization and fairness in channel allocation for cognitive radio networks. Each objective function denotes a secondary user's throughput. The secondary users attempt to achieve an agreement across each other that no one is eager to change its situation. Here, there is a set of solutions, namely Pareto optimal set. This paper proposes using ε-constraint method to obtain this solution set. The network was simulated for different numbers of secondary users, primary users, and channels. The corresponding Pareto set was obtained via ε-constraint method. The simulation outcome justifies the effectiveness of the method to attain Pareto sets in an acceptable time. The proposed method can be applied to applications such as wireless sensor networks and autonomous cars.

As the future work, it is a good to determine the optimal number of GPs for different topologies. In addition, it is possible to consider multiple CR networks rather than one network, to formulate several multi-objective problems that each of them is for one CR network. Moreover, the paper assumes that environmental conditions such as available spectrum and user location are static during the allocation period. Thus, it is as a good next step if spectrum channels and user location dynamically change during the allocation period.

## APPENDIX I

Pseudo code for network modeling [8]

Deploy $K$ PUs: each PU $k$ $(1 \leq k \leq K)$ locates in $x_k$, and uses the channel $y_k$.
Deploy $N$ SUs: each SU $n$ $(1 \leq n \leq N)$ locates in $\phi_n$.

```
For n = 1 to N do
    d_s(n,m) = min(d_max, min_{k,y_k=m} {dist(φ_n, x_k) − d_{p_k}})
    If d_s(n,m) > d_min
        [B]_{n,m} = d_s(n,m)^2,  [L]_{n,m} = 1
    Else
        [B]_{n,m} = [L]_{n,m} = 0
    End if
End for
For n = 1 to N − 1 do
    For i = n + 1 to N do
        For m = 1 to M do
            If d_s(n,m) + d_s(i,m) ≥ dist(φ_n, φ_i)
                [C]_{n,i,m} = [C]_{i,n,m} = 1
            Else
                [C]_{n,i,m} = [C]_{i,n,m} = 0
            End if
        End for
    End for
End for
```

A description of Appendix I

The position of secondary users $(1 \leq n \leq N)$ and primary users $(1 \leq k \leq K)$ are randomly selected within a wireless network area. This code includes two parts: (1) Entries of **B** and **L** are constructed in which the interference range of the $n^{th}$ SU on the $m^{th}$ channel $d_s(n,m)$ is equal to minimum $Dist(n,x) - d_p(x,m)$ and $d_{\max}$; (2) Elements of **C** are calculated. Here, two SUs would interfere if their relative distance is less than the sum of their interference ranges.

## REFERENCES

[1] FCC Spectrum Policy Task Force. Report of the spectrum efficiency working group. Technical Report, http://www.fcc.gov/sptf/reports.html, 2002.

[2] F. Khoseimeh and S. Haykin, "Dynamic spectrum management for cognitive radio: an overview," Wirel. Commun. Mob. Comput., Vol. 9, pp. 1447-1459, Jan. 2009.

[3] I. F. Akyildiz, W. Lee, M. C. Vuran, and S. Mohanty, "Next Generation/Dynamic Spectrum Access/Cognitive Radio Wireless Networks: A Survey," Comp. Networks J., Vol. 50, pp. 2127-59, Sept. 2006.

[4] N. Nie, C. Comaniciu, "Adaptive channel allocation spectrum etiquette for cognitive radio networks," Proc. IEEE DySPAN, pp. 269-278, 2005.

[5] J. Huang, R. Berry, and M. L. Honig, "Auction-based spectrum sharing," ACM/Springer Mobile Networks Apps., Vol. 11, No. 3, pp. 405-418, June 2006.

[6] C. Kloeck, H. Jaekel, and F. K. Jondral, "Dynamic and local combined pricing, allocation and billing system with cognitive radios," Proc. IEEE DySPAN, pp. 73-81, 2005.

[7] H. Zheng and C. Peng, "Collaboration and fairness in opportunistic spectrum access," Proc. 40th Annu. IEEE Int. Conf. Communications (ICC), pp. 3132-3136, 2005.

[8] C. Peng, H. Zheng, and B. Y. Zhao, "Utilization and fairness in spectrum assignment for opportunistic spectrum access," ACM Mobile Networks and Applications (MONET), Vol. 11, No. 4, pp. 555-576, 2006.

[9] D. Niyato and E. Hossain, "Competitive spectrum sharing in cognitive radio networks: A dynamic game approach," IEEE Trans. Wireless Comm., vol. 7, No. 7, pp. 2651-2660, Jul. 2008.

[10] F. Wang, M. Krunz, and S. Cui, "Price-based spectrum management in cognitive radio networks," IEEE J. Sel. Topics Signal Process., Vol. 2, No. 1, pp. 74-87, Feb. 2008.

[11] Z. Han, Z. Ji and K.R. Liu, "Low-complexity OFDMA channel allocation with Nash bargaining solution fairness," Proc. of Globecom, Dallas, Texas, Nov. 2004.

[12] Z. Ji and K. J. Ray Liu, "Belief-assisted pricing for dynamic spectrum allocation in wireless networks with selfish users," Proc. IEEE SECON, pp. 119-127, Jan. 2006.

[13] G. Salami, O. Durowoju, A. Attar, O. Holland, R. Tafazolli, and H. Aghvami, "A Comparison Between the Centralized and Distributed Approaches for Spectrum Management," IEEE Commun. Sur. Tutor., pp.1-17. 2010. DOI: 10.1109/SURV.2011.041110.00018,

[14] A. Ghasemi, M. A. Masnadi-Shirazi, M. Biguesh, F. Qassemi, "Spectrum Assignment Based on Bee Algorithms in Multi-Hop Cognitive Radio Networks," IET Communication Journal, Vol. 8, Issue: 13, pp. 2356-2365, 05 Sep. 2014.

[15] A. Ghasemi, M. Masnadi Shirazi, M. Biguesh, F. Qassemi, "Spectrum Allocation with Control of Interference Based on Differential Evolution Algorithm between Cognitive Radio Users," ICEE2012 (20th Iranian Conference on Electrical Engineering (ICEE)), Tehran, Iran, April, 2012.

[16] A. Ghasemi, A. F. Jahromi, M. A. Masnadi-Shirazi, M. Biguesh, F. Ghasemi, "Spectrum Allocation Based on Artificial Bee Colony in Cognitive Radio Networks," 6th International Symposium on Telecommunication (IST), pp. 182-187, Tehran, Iran, 6-8 Nov. 2012.

[17] G. N. Courant, "Classic Methods for Multi-Objective Optimization," Institute of Mathematical Sciences, New York University, 31 Jan. 2008.

[18] G. Mavrotas, "Effective implementation of the ε-constraint method in multi-objective mathematical programming

problems modified augmented," Appl. Math. and Comp. Vol. 213, pp. 455–465, 2009.

[19] Generalized Algebraic Modeling Systems (GAMS). <http://www.gams.com>.

[20] K. Deb, A. Pratap, S. Agarwal, and T. Meyarivan, "A fast and elitist multiobjective genetic algorithm: NSGA-II," IEEE Trans. Evolutionary Comp., Vol. 6, Issue: 2, pp. 182-197, Apr. 2002.

[21] C.A. Coello Coello and M.S. Lechuga, "MOPSO: a proposal for multiple objective particle swarm optimization," Proceedings of the 2002 Congress on Evolutionary Computation. CEC'02, Honolulu, HI, USA, USA, 12-17 May 2002.

[22] G. Guillén-Gosálbez, "A novel MILP-based objective reduction method for multi-objective optimization: application to environmental problems," Computer and Chemical Eng., Vol. 35, Issue: 8, pp. 1469-1477, 10 Aug. 2011.